\documentclass[journal]{IEEEtran}
\usepackage{amsmath,amsfonts}
\usepackage{algorithmic}
\usepackage{algorithm}
\usepackage{array}
\usepackage[caption=false,font=normalsize,labelfont=sf,textfont=sf]{subfig}
\usepackage{textcomp}
\usepackage{stfloats}
\usepackage{url}
\usepackage{verbatim}
\usepackage{multirow}
\usepackage{colortbl, makecell}
\usepackage{graphicx}
\usepackage{hyperref}
\usepackage{cite}
\usepackage{pifont}% http://ctan.org/pkg/pifont
\newcommand{\cmark}{\ding{51}}%
\newcommand{\xmark}{\ding{55}}%
\usepackage{tabularx}
\usepackage[commandnameprefix=ifneeded,defaultcolor=black]{changes}
\begin{document}

\author{Bang Zeng, Student Member, IEEE, Beilong Tang, Wang Xiang, Ming Li\IEEEauthorrefmark{1}, Senior Member, IEEE
\IEEEcompsocitemizethanks{
	\IEEEcompsocthanksitem Bang~Zeng, Wang Xiang and Ming~Li are with the School of Computer Science, Wuhan University, Wuhan 430072, China, and also with Suzhou Municipal Key Laboratory of Multimodal Intelligent Systems, Digital Innovation Research Center, Duke Kunshan University, Kunshan 215316, China. Beilong Tang is with the North Carolina State University (e-mail: bangzeng@whu.edu.cn; btang5@ncsu.edu; 2025102110031@whu.edu.cn; ming.li369@dukekunshan.edu.cn).}
\thanks{\IEEEauthorrefmark{1} Corresponding author.}}
        % <-this % stops a space


\title{Discriminative–Generative Target Speaker Extraction with Decoder-Only Language Models}
\maketitle

\begin{abstract}
Target speaker extraction (TSE) aims to recover the speech of a desired speaker from a mixture given a short enrollment utterance, while speech enhancement (SE) focuses on improving speech quality under noisy conditions. Most existing TSE and SE systems are based on discriminative modeling and have shown strong interference suppression ability, but they often remain limited in perceptual quality and naturalness. To address this issue, we first introduce LauraTSE, a generative TSE model built on an autoregressive decoder-only language model. Although generative modeling is promising for quality enhancement, purely generative TSE may suffer from hallucination, content drift, and limited controllability in complex acoustic conditions. We therefore propose a discriminative–generative two-stage framework, where a discriminative front-end first produces target-related representations with strong interference suppression, and a generative back-end then reconstructs high-quality speech in the neural audio codec representation space. This design combines the controllability of discriminative extraction with the reconstruction capability of generative modeling. We further investigate several collaboration strategies for the two-stage framework, including front-end freezing, joint fine-tuning, SI-SDR regularization, and autoregressive/non-autoregressive inference. Experimental results on both TSE and SE benchmarks show that the proposed framework achieves a better balance among perceptual quality, intelligibility, and speaker consistency than purely discriminative or purely generative baselines.
\end{abstract}


\section{Introduction}
\IEEEPARstart{H}{umans} can selectively attend to a target speech stream in complex acoustic environments, a phenomenon commonly referred to as the cocktail party effect \cite{Cherry1953SomeEO, bronkhorst2000cocktail}. Inspired by this ability, speech separation has been extensively studied for decades. Early approaches, such as non-negative matrix factorization (NMF) ~\cite{schmidt2006single} and computational auditory scene analysis (CASA)~\cite{lyon1983computational}, mainly relied on spectro-temporal masking and often degraded under acoustically challenging conditions. With the development of deep learning, neural speech separation methods, including deep clustering~\cite{hershey2016deep}, deep attractor networks (DANet)\cite{chen2017deep}, and permutation invariant training (PIT)\cite{yu2017permutation}, substantially improved separation performance. Time-domain models such as TasNet~\cite{luo2018tasnet} and its variants further advanced waveform reconstruction quality by avoiding explicit phase estimation. More recent architectures in both the time and time-frequency domains have continued to improve separation accuracy.  However, most conventional speech separation systems~\cite{luo2020dual,tzinis2020sudo,luo2019conv,zeghidour2021wavesplit,li2021dual,chen2020dual,subakan2021attention,li2022efficient,zhao2023mossformer,zhao2024mossformer2} aim to recover all speakers in a mixture and often assume that the number of sources is known in advance, which limits their practicality in real-world scenarios.

In contrast, target speaker extraction (TSE)~\cite{wang19h_interspeech,vzmolikova2019speakerbeam,hao2020unified,li20p_interspeech,ge2020spex+,wang2021neural,ge2021multi,liu2023x,hao2024x,elminshawi2024new} focuses on extracting only the desired speaker from a mixture with the help of an enrollment utterance. This formulation is especially attractive in realistic applications, where the number of interfering speakers may be unknown and only one target speaker is of interest. %As illustrated in Fig.~\ref{fig:tytse}, 
A typical TSE system follows an encoder–separator–decoder architecture in either the time domain or the time-frequency domain. Most existing TSE methods rely on a speaker embedding extractor to derive a compact representation from the enrollment speech and then use this representation to guide target extraction. However, speaker embedding extractors are usually optimized for speaker recognition rather than TSE. As a result, they may discard fine-grained acoustic details that are useful for accurate target reconstruction. To mitigate this mismatch, speaker-embedding-free TSE methods~\cite{zeng2023sef,hu2024smma,yang2024target,11012711} have been proposed to exploit reference speech representations more directly.

Despite these advances, most existing TSE methods still adopt a discriminative paradigm that directly learns a deterministic mapping from the mixture and enrollment speech to the target signal. Such models are generally effective at suppressing interferers and preserving target controllability. However, they are typically optimized with signal-level objectives that do not always align well with human auditory perception, and they may struggle to recover fine-grained details that are missing or distorted during extraction~\cite{distortion}. As a result, discriminative TSE systems often remain limited in perceptual naturalness and reconstruction fidelity, especially under challenging acoustic conditions. 

Generative models offer a different perspective by modeling speech distributions rather than a single deterministic solution. In principle, this allows them to better reconstruct plausible speech details and improve perceptual quality~\cite{target_diff,tokensplit,zhang2025anyenhance,kang2025llase}. Recent studies have explored several generative paradigms for TSE, including diffusion models~\cite{target_diff}, variational autoencoders (VAEs)~\cite{vae}, and language-model-based approaches~\cite{zhang2025anyenhance, tang2024tselm}. These works provide encouraging evidence that generative modeling can benefit TSE. Nevertheless, autoregressive decoder-only language models remain underexplored in this task.  Although SpeechX~\cite{wang2024speechx} demonstrates the potential of decoder-only language models in multi-task speech processing, it does not directly answer a key question for TSE: can a compact decoder-only language model provide sufficient modeling capacity for high-quality target speaker extraction in a task-specific setting?

% %
% \begin{figure*}[t!]
%   \centering
%   % \captionsetup{skip=2pt}
%   %\hspace*{0.5cm}
%   % \setlength{\abovecaptionskip}{-0.8cm}
%   \includegraphics[width=0.75\linewidth]{fig/typical_tse.pdf}
%   \caption{The diagram of a typical target speaker extraction method. The speaker embedding extractor is typically a pre-trained speaker recognition model. 'C' denotes the concatenation.}
%   \label{fig:tytse}
%   % \vspace{-0.6cm}
% \end{figure*}
% %

To investigate this question, we previously introduced LauraTSE~\cite{tang2025lauratse}, a generative TSE model based on an autoregressive decoder-only language model. LauraTSE predicts coarse codec representations of the target speech conditioned on continuous representations of the mixture and enrollment speech, and then refines them with an encoder-only module to recover fine-grained acoustic details. Although LauraTSE improves perceptual quality, purely generative TSE still faces important limitations. In particular, generative reconstruction may suffer from hallucination, content drift, and reduced controllability when the conditioning information is imperfect or insufficient. These limitations raise concerns about the robustness and reliability of purely generative TSE systems in complex acoustic scenarios.

Motivated by these observations, this paper proposes a discriminative–generative two-stage framework for TSE. The key idea is to decompose the task into two complementary stages. A discriminative front-end first performs target-oriented extraction and interference suppression, producing structured intermediate representations that retain strong target relevance. A generative back-end then reconstructs high-quality target speech from these intermediate representations in the neural audio codec space. In this way, the front-end reduces the burden of coarse target localization, while the back-end focuses on detail reconstruction and perceptual refinement. The resulting framework aims to improve perceptual quality without sacrificing intelligibility and speaker consistency.

This work substantially extends our preliminary LauraTSE study~\cite{tang2025lauratse}. Beyond the original purely generative model, we introduce a hybrid discriminative–generative framework for both target speaker extraction (TSE) and speech enhancement (SE), and instantiate it as USEF-Laura-TSE by combining USEF-TFGridNet~\cite{11012711} with LauraTSE, and as BSRNN-Laura-SE by integrating BSRNN~\cite{10096020} with the same generative back-end. We further provide a systematic analysis of front-end freezing versus joint fine-tuning, SI-SDR regularization, and autoregressive/non-autoregressive inference. These analyses reveal how discriminative controllability and generative flexibility interact in TSE and SE, and how different training and inference strategies affect the trade-off among perceptual quality, intelligibility, and speaker consistency.

The main contributions of this article are summarized as follows:
\begin{itemize}
\item We develop LauraTSE, a generative TSE model based on an autoregressive decoder-only language model. LauraTSE bridges continuous acoustic conditioning and neural audio codec representations, enabling task-specific generative modeling for TSE without relying on explicit speaker embeddings.

\item We propose a discriminative–generative two-stage framework for both target speaker extraction and speech enhancement, instantiated as USEF-Laura-TSE and BSRNN-Laura-SE, respectively. In the proposed framework, a discriminative front-end first produces target-aligned and interference-suppressed intermediate representations, which are then refined by a generative back-end to enhance perceptual quality.

\item We present a systematic study of collaboration strategies between the two stages. Specifically, we analyze front-end freezing versus joint fine-tuning, SI-SDR regularization, and autoregressive/non-autoregressive inference, showing how these design choices affect the trade-off among perceptual quality, intelligibility, and speaker consistency.
\end{itemize}

\section{Related Works}
\subsection{Discriminative Approaches for Target Speaker Extraction}

Discriminative TSE methods have achieved substantial progress in recent years and can be broadly divided into time-domain and time-frequency-domain approaches. Early TSE systems mainly estimated speaker-dependent masks on short-time Fourier transform (STFT) features. Time-domain methods later gained attention because they avoid explicit phase reconstruction and often yield better waveform quality. Representative models such as TasNet~\cite{luo2018tasnet} and Conv-TasNet~\cite{luo2019conv} employ convolutional encoder--decoder structures to learn waveform-level representations. More advanced architectures, including DPRNN~\cite{luo2020dual}, SepFormer~\cite{subakan2021attention}, and transformer-based models~\cite{chen2020dual}, further enhance extraction performance by explicitly modeling long-range temporal dependencies and global contextual information. For the TSE task, target speaker information is typically incorporated through speaker embeddings extracted from reference speech, which are used to guide the extraction process~\cite{wang19h_interspeech,vzmolikova2019speakerbeam}. In such embedding-based frameworks, speaker encoders~\cite{he2016deep,chung2020in,wan2018generalized,deng2019arcface} are integrated with separation networks via feature concatenation, conditioning, or attention mechanisms, and various architectural designs have been proposed to improve robustness in feature extraction and cross-stream fusion. 

More recently, speaker-embedding-free TSE approaches~\cite{xiao2019single,yang2023target,zeng2023sef,hu2024smma,yang2024target} have been introduced to avoid the limitations of fixed-dimensional speaker embeddings. Instead of compressing the enrollment utterance into a single speaker vector, these methods directly exploit frame-level acoustic features and model the interaction between the enrollment and mixture streams using attention-based mechanisms. By preserving richer temporal and spectral details, such approaches can better alleviate the information loss and representation mismatch often caused by conventional speaker embeddings~\cite{snyder2018x,li2024tabe}.

Despite their strong target controllability and interference suppression abilities, most discriminative TSE systems are still trained with deterministic signal reconstruction objectives. This limits their ability to model the uncertainty and multimodality of speech signals and often makes it difficult to restore fine-grained acoustic details and naturalness. These limitations motivate the introduction of generative modeling into TSE.

\subsection{Generative Approaches for Target Speaker Extraction}
Generative TSE methods can be roughly categorized into continuous and discrete modeling paradigms. Continuous generative models, such as diffusion models~\cite{scheibler2023diffusion,richter2023speech,lu2022conditional,lemercier2023storm,kamo23_interspeech,zhang2024ddtse} and variational autoencoders (VAEs)~\cite{vae}, directly model the distribution of target speech and typically provide strong reconstruction fidelity. However, their computational cost and inference latency often limit their practicality, especially in real-time or resource-constrained scenarios. 

Recent work has increasingly explored discrete-representation-based generative methods built on neural audio codecs and large language models (LLMs)~\cite{devlin2019bert,brown2020language,raffel2020exploring}. In these systems, speech is first converted into discrete codec tokens, then generated conditionally with a language model, and finally reconstructed into waveforms by a codec decoder. Benefiting from strong sequence modeling capability, language-model-based methods have shown promise in speech enhancement, separation, and target speaker extraction. Among these architectures, decoder-only language models are particularly appealing because of their autoregressive formulation and flexibility for speech generation tasks~\cite{du2024funcodec,wang2024speechx,du2023lauragpt}.

Nevertheless, purely generative TSE also faces important challenges. Discrete token prediction can suffer from error accumulation, hallucination, and reduced stability, while large model sizes increase computational costs. More importantly, when the conditioning input does not sufficiently preserve target-related structure, purely generative reconstruction may fail to maintain semantic fidelity and speaker consistency. These observations motivate a hybrid design in which a discriminative front-end performs robust target-oriented extraction, and a generative back-end focuses on perceptual refinement. The present work follows this direction.

\section{Discriminative–Generative Target Speaker Extraction}
In this section, we first introduce LauraTSE in detail. We then present the proposed discriminative–generative two-stage framework, followed by a comprehensive description of its architecture and design principles. Finally, to validate the effectiveness of the two-stage framework, we construct a complete system, USEF-Laura-TSE, that employs USEF-TFGridNet~\cite{11012711} as the discriminative front-end and LauraTSE as the generative back-end.

\subsection{LauraTSE}
\label{sec:lauratse}
\begin{figure*}[t!]
  \centering
  % \captionsetup{skip=2pt}
  %\hspace*{0.5cm}
  % \setlength{\abovecaptionskip}{-0.8cm}
  \includegraphics[width=0.80\linewidth]{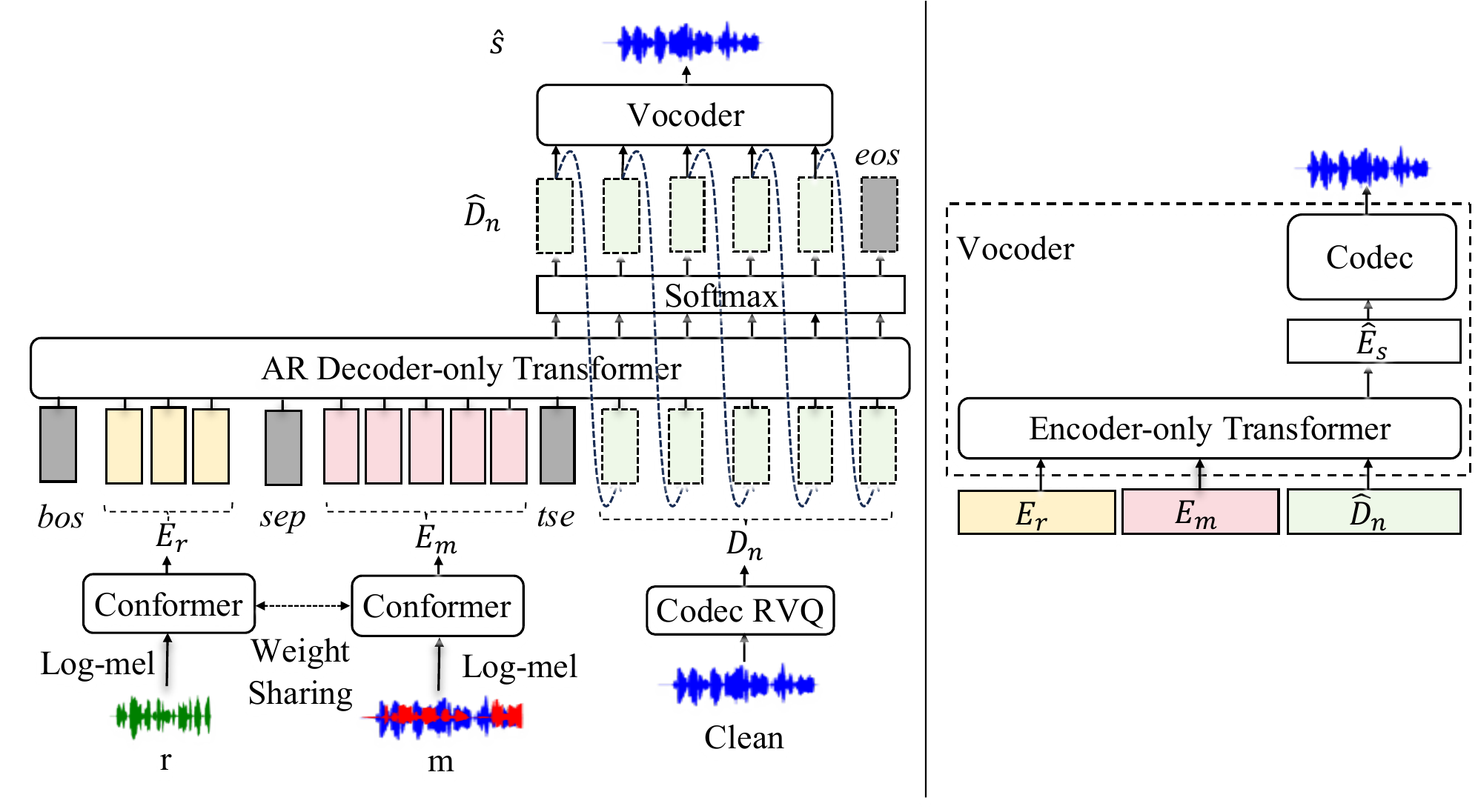}
  \caption{The diagram of LauraTSE network. ‘m’ and ‘r’ denote the mixed speech and reference speech, respectively. We use two weight sharing conformer to process the mixed and reference speech separately.}
  \label{fig:lauratse}
  % \vspace{-0.6cm}
\end{figure*}

In this study, we propose LauraTSE, a target-speaker extraction method based on an auto-regressive (AR) decoder-only language model built on the LauraGPT~\cite{du2023lauragpt} backbone. LauraTSE takes the log-mel spectrogram features of both the target speaker’s enrollment speech and the mixed speech as inputs, and employs the residual vector quantization (RVQ) layers of a neural audio codec to discretize audio representations, enabling high-quality modeling and reconstruction of the target speaker’s speech. The overall architecture of LauraTSE is illustrated in Fig~\ref{fig:lauratse}. LauraTSE consists of two key components. The first is an AR decoder-only language model that predicts the discrete representations of the target speech corresponding to the first several codec encoding layers. The second is a one-step encoder-only language model that jointly exploits information from the mixed and enrollment speech to directly predict the summed embeddings of all codec layers, thereby compensating for the limitations of auto-regressive modeling in modeling long-range temporal dependencies and mitigating error accumulation. In the following, we provide a detailed description of LauraTSE's architecture and design.

\subsubsection{Encoder}
The first stage of LauraTSE is the encoding stage. Following LauraGPT's processing strategy for speech enhancement, we first compute log-mel spectrogram features for both the enrollment speech and the mixed speech, denoted as \(\textbf{M}_\text{m}\) and \(\textbf{M}_\text{r}\). These two feature streams are then fed into a parameter-sharing Conformer~\cite{gulati2020conformer} encoder, producing continuous representations for the reference speech and the mixed speech:
\begin{equation}
  \textbf{E}_{\text{m}} = \text{C}(\textbf{M}_\text{m})
  \label{eq1}
\end{equation}
\begin{equation}
  \textbf{E}_{\text{r}} = \text{C}(\textbf{M}_\text{r})
  \label{eq2}
\end{equation}
where \(\textbf{E}_{\text{m}}\) \(\in\) \(\mathbb{R}^{\text{N} \times \text{L}_{m}}\) and \(\textbf{E}_{\text{r}}\) \(\in\) \(\mathbb{R}^{\text{N} \times \text{L}_{r}}\) represent the encoded outputs of the \(\textbf{M}_\text{m}\) and \(\textbf{M}_\text{r}\), respectively. \(\text{C}(\cdot)\) denotes the conformer block. N is the feature dimension. \added{\(\text{L}_\text{{m}}\)} and \added{\(\text{L}_\text{{r}}\)} are the number of time steps.

This encoding stage serves as a feature adapter within the overall framework. Rather than directly performing target speaker extraction, its primary objective is to map raw acoustic features into a continuous representation space that is more suitable for subsequent modeling by the auto-regressive decoder-only language model, thereby providing high-quality and structured inputs for generative modeling. It is worth noting that, unlike SpeechX\cite{wang2024speechx}, which uses discrete representations produced by a neural audio codec as inputs to the AR model, this work, like LauraGPT~\cite{du2023lauragpt}, preserves task-driven continuous feature representations. This design choice avoids potential information loss introduced by discretization, particularly for fine-grained speaker-related acoustic characteristics.

\subsubsection{Auto-Regressive Decoder-Only Language Model}
The auto-regressive decoder-only language model is designed to learn and predict the joint probability distribution of the coarse-grained discrete representations of the target speech, conditioned on the enrollment speech and the mixed speech. Specifically, the model factorizes the joint distribution of the target speech representations according to the chain rule of probability as follows:
\begin{equation}
	\textbf{P}_\theta(\hat{\textbf{D}}_\text{n}\mid \textbf{E}_{\text{m}},\textbf{E}_{\text{r}}) = \prod_{\text{i}\le \text{T}}{\textbf{P}_\theta(\hat{\textbf{D}}_\text{n}^{(i)}\mid \hat{\textbf{D}}_\text{n}^{(1:i-1)}, \textbf{E}_{\text{m}},\textbf{E}_{\text{r}})}
	\label{eq3}
  \end{equation}
where $T$ denotes the length of the output signal, and $\theta$ denotes the model parameters, and $\hat{\textbf{D}}_n$ denotes the generated discrete representation of the target speech.

During training, the input sequence to the AR decoder-only language model is organized as
$[\text{bos},\ \textbf{E}_\text{r},\ \text{sep},\ \textbf{E}_\text{m},\ \text{tse},\ \textbf{D}_\text{n}]$,
where \text{bos} is a learnable beginning-of-sequence token, \text{sep} separates the enrollment and mixed speech embeddings, \text{tse} marks the boundary between conditional inputs and target outputs, and $\textbf{D}_\text{n}$ denotes the sum of embeddings from the first n residual vector quantization (RVQ) layers of the target speech. The AR model is trained to predict the discrete representations of the first n RVQ layers. After generating hidden states, n parallel linear layers estimate token distributions for each RVQ layer, and a cross-entropy loss is applied between the predicted and ground-truth token distributions. The predicted tokens are then mapped to continuous embeddings using the codec decoder’s embedding tables and summed across layers to form a coarse-grained representation $\hat{\textbf{D}}_\text{n}$. During inference, the model generates $\hat{\textbf{D}}_\text{n}$ auto-regressively, frame by frame.

\subsubsection{Vocoder}
The objective of the vocoder module is to reconstruct a high-fidelity time-domain waveform of the target speaker by fully leveraging information from the mixed speech and the enrollment speech, based on the coarse-grained representations generated by the auto-regressive model. To this end, we design a vocoder consisting of an encoder-only language model and a frozen, pre-trained neural audio codec decoder. The encoder-only language model is built upon self-attention mechanisms, enabling effective modeling of long-range temporal dependencies and capturing fine-grained acoustic structures and speech details. Unlike SpeechX~\cite{wang2024speechx}, which predicts RVQ codes layer-wise, our design adopts a one-step encoder-only language model that directly predicts the summed embeddings across all RVQ layers of the target speech. This formulation substantially simplifies the modeling process while improving both training and inference efficiency.

Specifically, the input to the encoder-only language model is the concatenated feature sequence \([\textbf{E}_\text{r}, \textbf{E}_\text{m}, \hat{\textbf{D}}_\text{n}]\):
\begin{equation}
  [\,\cdot,\, \cdot,\, \hat{\textbf{E}}_\text{s}] = \text{EL}([\textbf{E}_\text{r}, \textbf{E}_\text{m}, \hat{\textbf{D}}_\text{n}])
  \label{eq4}
\end{equation}
where \(\textbf{E}_\text{r}\) and \(\textbf{E}_\text{m}\) denote the continuous embeddings of the enrollment speech and the mixed speech, respectively, and \(\hat{\textbf{D}}_\text{n}\) represents the embedding corresponding to the coarse-grained target speech representation generated by the first-stage AR decoder-only language model. \(\text{EL}(\cdot)\) denotes the encoder-only language model. The encoder-only model processes this sequence and outputs \([\,\cdot,\, \cdot,\, \hat{\textbf{E}}_\text{s}]\), where \(\hat{\textbf{E}}_\text{s}\) denotes the predicted fine-grained acoustic embedding of the target speaker. During training, the predicted embedding \(\hat{\textbf{E}}_\text{s}\) is supervised against the ground-truth target speech embedding \(\textbf{E}_\text{s}\), obtained from the neural audio codec as the sum of embeddings across all RVQ layers. Both L1 and L2 losses are jointly employed to optimize reconstruction accuracy and training stability. Finally, the frozen codec decoder converts the predicted embedding \(\hat{\textbf{E}}_\text{s}\) into the time-domain waveform of the target speaker’s speech. It is worth emphasizing that the AR decoder-only language model and the encoder-only language model are jointly trained end-to-end. 

\subsection{Architecture}
\begin{figure}[t!]
  \centering
  % \captionsetup{skip=2pt}
  %\hspace*{0.5cm}
  % \setlength{\abovecaptionskip}{-0.8cm}
  \includegraphics[width=0.8\linewidth]{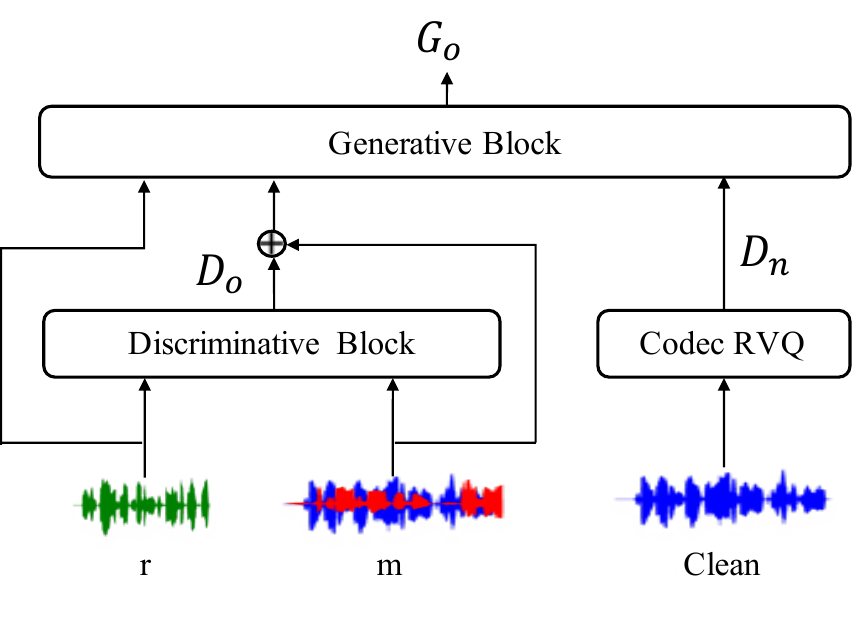}
  \caption{The diagram of discriminative-generative target speaker extraction framework. ‘m’ and ‘r’ denote the mixed speech and reference speech, respectively.}
  \label{fig:dgtse}
  % \vspace{-0.6cm}
\end{figure}

To leverage the advantages of both discriminative and generative approaches simultaneously, this work proposes a two-stage discriminative–generative framework for target speaker extraction. As illustrated in Fig.~\ref{fig:dgtse}, the framework consists of two collaborative modules: a discriminative module and a generative module. The discriminative module explicitly extracts target-speaker–related speech components or intermediate acoustic representations from the mixed speech, providing high-quality and low-interference conditional inputs for the subsequent generative module. The generative module then performs generative reconstruction based on the outputs of the discriminative module, further enhancing the perceptual quality of the target speech.

In the discriminative module, the discriminative block takes the reference speech \textbf{r} and the mixed speech \textbf{m} as inputs, and extracts target-related information by suppressing interference from non-target speakers. This module outputs a coarse target representation \(\textbf{D}_\text{o}\), which can be interpreted as an estimated target speech signal or an intermediate acoustic representation:
\begin{equation}
  \textbf{D}_\text{o} = \mathcal{D}(\textbf{m}, \textbf{r})
  \label{eq5}
\end{equation}
where \(\mathcal{D}(\cdot)\) denotes the discriminative extraction function.

In parallel, the ground-truth clean target speech is encoded by a neural audio codec with RVQ, producing a coarse discrete representation \(\textbf{D}_\text{n}\):
\begin{equation}
  \textbf{D}_\text{n} = \mathcal{Q}(\textbf{s})
  \label{eq6}
\end{equation}
where \textbf{s} denotes the clean target speech and \(\mathcal{Q}(\cdot)\) represents the codec encoder.

In the generative module, the generative block takes the discriminative output \(\textbf{D}_\text{o}\) as a conditional input and leverages generative modeling to reconstruct a refined target speech representation. The final output \(\textbf{G}_\text{o}\) is generated as:
\begin{equation}
  \textbf{G}_\text{o} = \mathcal{G}(\textbf{r}, \textbf{D}_\text{o}, \textbf{D}_\text{n})
  \label{eq7}
\end{equation}
where \(\mathcal{G}(\cdot)\) denotes the generative reconstruction function. At inference time, only \(\textbf{D}_\text{o}\) is required, and the generative block produces the enhanced target speech output \(\textbf{G}_\text{o}\).

Through this two-stage design, the discriminative block provides a low-interference and well-aligned target representation. In contrast, the generative block further refines speech details and improves perceptual quality via distribution-level modeling.

\subsection{Two-Stage Models to Target Speaker Extraction Tasks}
\begin{figure*}[t!]
  \centering
  % \captionsetup{skip=2pt}
  %\hspace*{0.5cm}
  % \setlength{\abovecaptionskip}{-0.8cm}
  \includegraphics[width=0.6\linewidth]{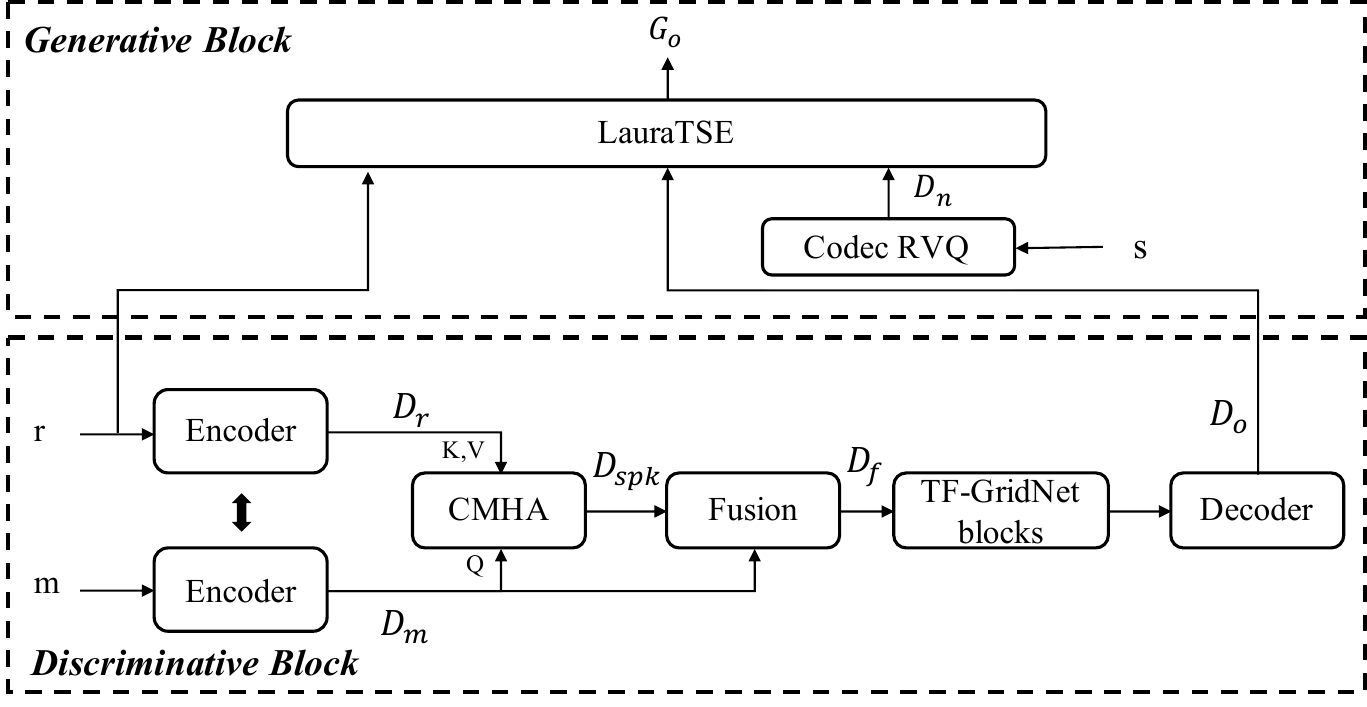}
  \caption{The diagram of USEF-Laura-TSE. ‘m’ and ‘r’ denote the mixed speech and reference speech, respectively.}
  \label{fig:usef-laura-tse}
  % \vspace{-0.6cm}
\end{figure*}

We construct a two-stage discriminative–generative target speaker extraction network, termed USEF-Laura-TSE, that employs USEF-TFGridNet~\cite{11012711} as the discriminative front-end and LauraTSE as the generative back-end. The overall architecture of USEF-Laura-TSE is illustrated in Fig~\ref{fig:usef-laura-tse}.

\subsubsection{Discriminative Block (USEF-TFGridNet)}
Given the reference speech \textbf{r} and the mixed speech \textbf{m}, both signals are first transformed into the time–frequency (T–F) domain using the short-time Fourier transform (STFT), followed by 2-D convolutional encoders:
\begin{equation}
  \textbf{\textbf{D}}_{\text{m}} = \text{Enc}(\textbf{m})
  \label{eq8}
\end{equation}
\begin{equation}
  \textbf{\textbf{D}}_{\text{r}} = \text{Enc}(\textbf{r})
  \label{eq9}
\end{equation}
where \(\mathrm{Enc}(\cdot)\) denotes the shared encoder composed of STFT and 2-D convolution layers.

\(\textbf{D}_{\text{m}}\) and \(\textbf{D}_{\text{r}}\) are fed into the CMHA module, where a cross multi-head attention mechanism is applied to extract frame-level features of the target speaker:
\begin{equation}
  \textbf{D}_{\text{spk}} = \text{CMHA}(\text{q}=\textbf{D}_{\text{m}}; \text{k},\text{v}=\textbf{D}_{\text{r}})
  \label{eq10}
\end{equation}
where \(\textbf{E}_{\text{m}}\) and \(\textbf{E}_{\text{r}}\) represent the encoder outputs of the mixed speech and reference speech, respectively. The Cross Multi-Head Attention operation is denoted as \(\text{CMHA}(\cdot)\), and \(\textbf{E}_{\text{spk}}\) is the output of the CMHA module. The CMHA module in USEF-TFGridNet~\cite{11012711} uses mixed speech encoding as the query. This approach produces a frame-level feature with the same length as \(\textbf{D}_{\text{m}}\), allowing the mixed and reference speech lengths to differ in the USEF-TFGridNet~\cite{11012711}.

The extracted speaker-aware representation \(\textbf{D}_{\text{spk}}\) is then fused by direct concatenation with the mixed-speech features:
\begin{equation}
  \textbf{D}_\text{f} = \text{Concat}(\textbf{D}_\text{m}, \textbf{D}_\text{spk})
  \label{eq11}
\end{equation}

The fused features are processed by a stack of TF-GridNet blocks to model global T–F dependencies. Finally, a decoder composed of 2-D transposed convolutions and inverse STFT (iSTFT) reconstructs the discriminative output \(\textbf{D}_{\text{o}}\).

\subsubsection{Generative Block (LauraTSE)}
The output of the discriminative block \(\textbf{D}_{\text{o}}\) is fed into the generative block as a conditional input. During training, the clean target speech s is encoded by a neural audio codec with RVQ to obtain a coarse discrete representation \(\textbf{D}_{\text{n}}\). LauraTSE learns to model the conditional distribution of the target speech and generates the final output \(\textbf{G}_{\text{o}}\). The detailed procedure of LauraTSE is described in Section~\ref{sec:lauratse}

\subsection{Two-Stage Models to Speech Enhancement Tasks}
To further validate the effectiveness of the proposed two-stage framework, we additionally construct a two-stage model for the speech enhancement task, termed BSRNN-Laura-SE. The discriminative front-end of BSRNN-Laura-SE adopts BSRNN~\cite{10096020}, while the generative back-end adopts the same generative architecture as LauraTSE. Its overall architecture is identical to that of USEF-Laura-TSE, except that the enrollment utterance is not required as input. For the neural audio codec, we consider two configurations: FunCodec~\cite{du2024funcodec} for 16-kHz audio and FlowDec~\cite{welker2025flowdec} for 48-kHz audio.

\section{Experimental Setup}
\subsection{Datasets}
\subsubsection{USEF-Laura-TSE}
The main experiments in this work are conducted using the 460-hour clean speech subset of the LibriSpeech~\cite{librispeech} corpus, referred to as LibriSpeech-460h.Training mixtures are generated online by randomly selecting and mixing clean utterances during training. The target-to-interference ratio is randomly sampled from 0 to 5 dB to simulate realistic TSE conditions. For validation, the clean development set of Libri2Mix~\cite{librimix} is used. During both training and evaluation, the enrollment speech is randomly cropped to 5 s to improve robustness to enrollment-duration variations. During testing, the clean test set of Libri2Mix is used, and an enrollment utterance is randomly selected for each target speaker to better reflect practical target-speaker extraction scenarios.

LauraTSE is first pre-trained on LibriSpeech-460h to learn robust speech and speaker representations from large-scale clean speech data, and is then fine-tuned on the Libri2Mix clean training set to adapt to mixture conditions. For the LauraTSE ablation studies, however, the model is trained only on the Libri2Mix clean training set to ensure controlled and fair comparisons. 

\subsubsection{BSRNN-Laura-SE}
For BSRNN-Laura-SE, we train the model using the official training data provided by the URGENT Challenge~\cite{saijo25_interspeech}. For evaluation, we use the official URGENT Challenge validation dataset.

\subsection{Network Configuration}
\subsubsection{LauraTSE}
For LauraTSE, we adopt LauraGPT~\cite{du2023lauragpt} as the backbone of the AR decoder-only language model and employ FunCodec~\cite{du2024funcodec} as the neural audio codec. The decoder-only model predicts the first n=2 codec layers, yielding a coarse target representation. 

For acoustic conditioning, both the enrollment speech and the mixture speech are converted into log-Mel features using a 512-sample analysis window and a 256-sample frame shift. These features are processed by a shared Conformer encoder with six layers, eight attention heads, and a hidden dimension of 512 to obtain continuous acoustic representations. The autoregressive decoder-only Transformer contains 10 Transformer blocks, each with eight attention heads and a hidden dimension of 512. Conditioned on the encoded enrollment and mixture representations, it predicts the discrete codec representations of the first n RVQ layers in a frame-by-frame autoregressive manner.

To reconstruct high-fidelity waveforms, we employ an encoder-only Transformer as a refinement module. This module contains six Transformer layers with eight attention heads and a hidden dimension of 512. It jointly exploits the mixture representation, the enrollment representation, and the autoregressively predicted codec representation to estimate the full RVQ embedding of the target speech for waveform reconstruction.

\subsubsection{USEF-Laura-TSE}
USEF-Laura-TSE employs USEF-TFGridNet~\cite{11012711} as the discriminative front-end. as the discriminative front-end. In the encoder, the input speech is transformed into the time-frequency domain using STFT and then processed by 2-D convolutional layers. The encoder output channels are set to 128. The cross-attention module uses a single-layer structure with four attention heads and a feed-forward hidden dimension of 512. In both the full-band and sub-band branches, BLSTM layers with 256 hidden units are used for sequence modeling. A cross-frame self-attention module with one attention layer, four heads, and a 512-dimensional feed-forward network is then applied to model global time-frequency dependencies.

The number of TF-GridNet blocks is set to 2 and 6 for USEF-TFGridNet-S and USEF-TFGridNet-L, respectively. In the decoder, 2-D transposed convolutions are used to project the encoded features back to the complex spectrum, which is then converted to the waveform domain. The discriminative output is finally fed to LauraTSE as the conditional input of the generative back-end.

\subsubsection{BSRNN-Laura-SE}
The BSRNN-Laura-SE model adopts BSRNN~\cite{10096020} as the front-end, with parameter settings kept consistent with the baseline model used in the URGENT Challenge~\cite{saijo25_interspeech}. The back-end employs the same generative architecture as LauraTSE, and its parameter settings are also kept identical.

\subsection{Training Details}
We adopt a stage-wise training strategy for the proposed discriminative–generative framework. The discriminative front-end is first trained independently on Libri2Mix to obtain stable target-oriented extraction behavior. After this pre-training stage, we consider three collaboration strategies when training the full framework.

First, the discriminative front-end can be frozen while only the generative back-end is optimized. This setting preserves the standalone behavior of the front-end and isolates the contribution of the back-end. Second, the front-end can be unfrozen and jointly optimized with the back-end, allowing the two modules to adapt to each other during end-to-end training. Third, an additional SI-SDR loss~\cite{le2019sdr} can be imposed on the output of the discriminative front-end to regularize its reconstruction behavior during joint optimization.

The SI-SDR loss is defined as follows:
\begin{equation}
\begin{cases}
 \textbf{s}_{\text{T}} = \frac{<\hat{\textbf{s}},\textbf{s}>\textbf{s}}{||\textbf{s}||^2}\\ 
\textbf{s}_{\text{E}} = \hat{\textbf{s}} - \textbf{s}_{\text{T}} \\
\text{SI-SDR} = -10\lg{\frac{||\textbf{s}_{\text{T}}||^2}{||\textbf{s}_{\text{E}}||^2}}
\end{cases}
\label{eq12}
\end{equation}
where \(\hat{\textbf{s}}\) \(\in\) \(\mathbb{R}^{1 \times \text{T}}\) represents the estimated target speaker speech, while \(\textbf{s}\) \(\in\) \(\mathbb{R}^{1 \times \text{T}}\) represents the clean source speech. In training, we minimize the negative SI-SDR.

The generative back-end is trained from scratch and does not rely on pre-trained weights. The overall LauraTSE model contains about 77M parameters, among which about 36M belong to the decoder-only Transformer. We use the Adam optimizer~\cite{KingBa15} with an initial learning rate of \(1 \times 10^{-3}\). A 10k-step warm-up schedule is applied for stabilization. When the validation performance does not improve for three consecutive epochs, the learning rate is halved. All models are trained for 100 epochs.

\begin{table}
  \renewcommand{\arraystretch}{1.2}
  \begin{center}
      \caption{
  Evaluation results for different Decoder-Encoder configurations. \textit{Decoder-Encoder-joint} and \textit{Decoder-Encoder-split} refer to the two integration strategies. \textit{Target-$n$} denotes the reconstructed target clean audio using only the first $n$ layers of the codec. \textit{No-Encoder} uses summation of only the first $n$ layers of the decoder-only LM output to generate speech without the encoder.
  }
    \begingroup
    \setlength{\tabcolsep}{6pt}  % Reduce column spacing only here
    \begin{tabular}{cccc}
      \Xhline{2\arrayrulewidth}
      Model & NISQA $\uparrow$ & dWER $\downarrow$ & WeSpeaker Sim $\uparrow$ \\
      \hline
      Decoder-Encoder-joint    & 4.241 & 0.241 & 0.847 \\
      Decoder-Encoder-split    & 4.253 & 0.232  & 0.858 \\
      \hline
      Target-$n$ ($n=2$)              & 3.644 & 0.301 & 0.740 \\
      No-Encoder               & 3.807 & 0.579 & 0.709 \\
      \Xhline{2\arrayrulewidth}
    \end{tabular}
    \endgroup
    \label{tbl:encoder_split}
  \end{center}
\end{table}
\begin{table}
  \caption{Input composition results for the encoder-only LM.}
  \renewcommand{\arraystretch}{1.2}
  \begin{center}
    \setlength{\tabcolsep}{5pt}
    \begin{tabular}{cccccccc}
      \Xhline{2\arrayrulewidth}
      \multirow{2}{*}{Model} & \multicolumn{3}{c}{Input} & \multirow{2}{*}{NISQA$\uparrow$} & \multirow{2}{*}{dWER $\downarrow$} & \multicolumn{1}{c}{WeSpeaker} \\
      & $E_r$ & $E_m$ & $D_n$ & & & Sim$\uparrow$ \\
      \hline
      Encoder-All   & \cmark & \cmark & \cmark & 4.241 & 0.241 & 0.847 \\
      Encoder-Mix      & \xmark & \cmark & \cmark   & 4.173 & 0.239 & 0.842 \\
      Encoder-Ref      & \cmark & \xmark & \cmark  & 4.187 & 0.480 & 0.763 \\
      \Xhline{2\arrayrulewidth}
    \end{tabular}
    \label{tbl:encoder_input}
  \end{center}
  % \vspace{-10pt}
\end{table}

\subsection{Evaluation Metrics}
\subsubsection{USEF-Laura-TSE}
Because codec-based waveform generation may introduce temporal and phase deviations relative to the clean reference, conventional intrusive measures may not fully reflect the perceptual quality of generative outputs.  Therefore, for USEF-Laura-TSE, intrusive metrics such as PESQ~\cite{rix2001perceptual} and STOI~\cite{taal2010short} are not adopted. Instead, we primarily employ the following evaluation metrics that are more suitable for generative speech modeling, most of which are non-intrusive. 

DNSMOS~\cite{dnsmos} is used as a non-intrusive quality metric and provides SIG, BAK, and OVRL scores ranging from 1 to 5. NISQA~\cite{mittag2021nisqa} is another non-intrusive perceptual metric that predicts an overall quality score from 1 to 5. SpeechBERT~\cite{saeki2024speechbertscore} is used to measure semantic similarity between the generated speech and the target speech in a self-supervised representation space, where HuBERT-base~\cite{hubert} is used for feature extraction. Differential Word Error Rate (dWER)~\cite{dwer} is used as an intelligibility-oriented metric and is computed with the Whisper-base ASR model~\cite{whisper}. Finally, speaker similarity is evaluated by cosine similarity in a speaker-embedding space using both WavLM-base\footnote{\href{https://huggingface.co/microsoft/wavlm-base-plus-sv}{https://huggingface.co/microsoft/wavlm-base-plus-sv}} and the \textit{ResNet\_221LM} and the ResNet 221LM model from WeSpeaker~\cite{wespeaker}.

\subsubsection{BSRNN-Laura-SE}
For BSRNN-Laura-SE, we use the evaluation toolkit and metrics adopted in the URGENT Challenge~\cite{saijo25_interspeech}, including DNSMOS~\cite{dnsmos}, NISQA~\cite{mittag2021nisqa}, ScoreQ~\cite{mittag2021nisqa}, PESQ~\cite{rix2001perceptual}, STOI~\cite{taal2010short}, speaker similarity (SpkSIM), and Character accuracy (cAcc = 1 - CER).

\begin{figure}
\centering
\includegraphics[width=0.8\linewidth]{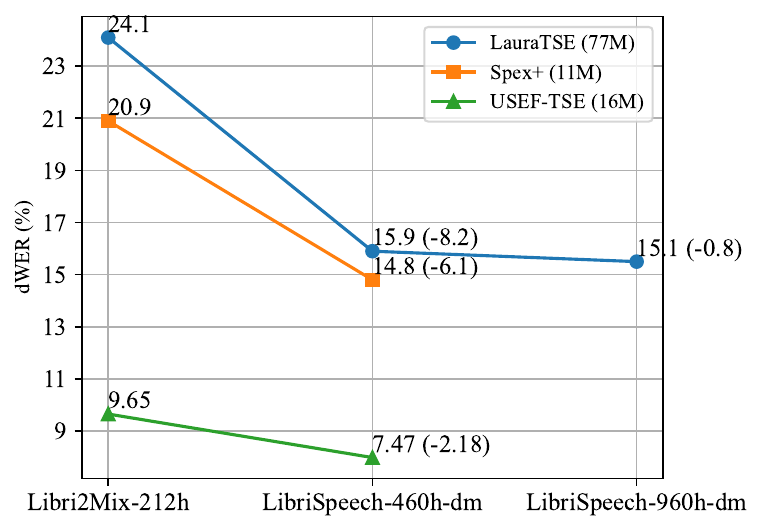}
\caption{dWER versus training data scale across models. Annotations "(-X)" denote relative dWER reduction (percentage points) compared to the preceding smaller dataset.}
\label{fig:data_scale}
% \vspace{-10pt}
\end{figure}
\section{Results and Discussions}
This section presents the experimental results of the proposed framework on LibriMix. We first analyze LauraTSE to understand the strengths and limitations of the decoder-only generative paradigm for TSE, and then evaluate USEF-Laura-TSE to examine how a discriminative front-end affects perceptual quality, intelligibility, and speaker consistency.

\begin{table*}
  \caption{Ablation studies of LauraTSE. \textit{n-} denotes the output layer number of the decoder-only LM. The \textit{Ref output} formats the output of the decoder-only LM to contain both the clean and reference speech. \textit{Discrete IO} uses discrete codec embeddings rather than continuous features as the input features. For \textit{WavLM input}, the WavLM~\cite{wavlm} embeddings are utilized as the input features.}
  \renewcommand{\arraystretch}{1.2}
  \begin{center}
    \setlength{\tabcolsep}{8pt}
    \begin{tabular}{cccccccccc}
      \Xhline{2\arrayrulewidth}
      \multirow{2}{*}{Model} & \multicolumn{3}{c}{DNSMOS $\uparrow$} & \multirow{2}{*}{NISQA $\uparrow$} & \multirow{2}{*}{SpeechBERT $\uparrow$} & \multirow{2}{*}{dWER $\downarrow$} & \multirow{2}{*}{WavLM Sim $\uparrow$} & \multirow{2}{*}{Wespeaker Sim $\uparrow$} \\
      \cline{2-4}
                             & SIG & BAK & OVL &                         &                      &                         &                          &                          \\
      \hline
      Base ($n$-2)          & 3.626 & 4.102 & 3.360 & 4.241 & 0.880 & 0.241 & 0.965  & 0.847 \\
      $n$-1   & 3.604 & 4.100 & 3.339 & 4.201 & 0.861 & 0.266 & 0.958  & 0.830 \\
      $n$-3    & 3.618 & 4.095 & 3.350 & 4.270 & 0.880 & 0.235 & 0.967  & 0.853 \\
      \hline
      Ref output    & 3.588 & 4.071 & 3.318 & 4.182 & 0.859 & 0.237 & 0.962 & 0.851 \\
      % \hline
      Discrete IO & 3.562 & 4.035 & 3.268 & 3.940 & 0.810 & 0.421 & 0.952  & 0.835 \\
      WavLM input  & 3.507 & 3.951 & 3.137 & 3.220 & 0.792 & 0.447 & 0.860 & 0.633        \\
      % \hline
      \Xhline{2\arrayrulewidth}
    \end{tabular}
    \label{tbl:ablation}
  \end{center}
  % \vspace{-10pt}
\end{table*}

\subsection{Ablation Results of LauraTSE}
\subsubsection{Data Scalability of LauraTSE}
To examine the scalability of LauraTSE with respect to training data size, we compare it with two representative discriminative TSE systems, SpEx+~\cite{ge2020spex+} and USEF-TSE (USEF-TFGridNet-L)~\cite{11012711}. SpEx+ and USEF-TSE contain about 11M and 16M parameters, respectively, whereas LauraTSE contains about 77M parameters. We train LauraTSE on three datasets with increasing scale: Libri2Mix-212h, LibriSpeech-460h-dm, and LibriSpeech-960h-dm.

% To evaluate the scalability of LauraTSE with respect to training data size, we compare it with two representative discriminative TSE models, namely SpEx+~\cite{ge2020spex+} and USEF-TSE (USEF-TFGridNet-L)~\cite{11012711}. The parameter sizes of SpEx+ and USEF-TFGridNet-L are approximately 11~M and 16~M, respectively, whereas LauraTSE contains approximately 77~M parameters. In this experiment, LauraTSE is trained using three progressively larger training datasets: 
% (1) \textit{Libri2Mix-212h}, derived from the Libri2Mix-clean subset and consisting of approximately 212 hours of clean mixed speech; 
% (2) \textit{LibriSpeech-460h-dm}, constructed from the LibriSpeech-460h subset using a dynamic mixing strategy; and 
% (3) \textit{LibriSpeech-960h-dm}, generated from the full 960-hour LibriSpeech dataset, also using dynamic mixing.

As shown in Fig.~\ref{fig:data_scale}, LauraTSE benefits substantially from increased training data. In particular, the dWER decreases from 24.1\% to 15.9\% and then to 15.1\% as the training data expand from Libri2Mix-212h to LibriSpeech-460h-dm and LibriSpeech-960h-dm, respectively. This trend is more pronounced than that of SpEx+, whereas USEF-TSE remains comparatively stable across different training scales. These results suggest that LauraTSE can exploit additional training data more effectively, likely because the generative model benefits from richer data diversity during distribution learning.

At the same time, the gain from 460 h to 960 h becomes relatively limited. This may be due to the increased difficulty of the train-other-500 portion of LibriSpeech-960h, as well as the finite modeling capacity of the current LauraTSE configuration. More importantly, although LauraTSE scales favorably with data, it does not consistently surpass strong discriminative baselines in semantic fidelity. This observation highlights that data scaling alone is not sufficient to resolve the reliability challenges of purely generative TSE.

\subsubsection{Effect of the Encoder-Only Refinement Module}

We next investigate the role of the encoder-only language model in LauraTSE. Since the autoregressive decoder-only model predicts only coarse codec representations, the encoder-only module is introduced to refine these outputs into higher-resolution continuous acoustic embeddings for waveform reconstruction.

We first compare joint training and split training, as shown in Table~\ref{tbl:encoder_split}. In the joint setting, the decoder-only and encoder-only modules are optimized end-to-end using the Straight-Through Estimator(STE)~\cite{ste}. In the split setting, the encoder-only module is trained separately using fixed outputs from the decoder-only model. The split training strategy yields slightly better performance overall, suggesting that strict end-to-end coupling between the two modules is not essential under the current configuration.

We further analyze the input composition of the encoder-only module. Table~\ref{tbl:encoder_input} compares three settings: Encoder-All, which uses the enrollment representation, the mixture representation, and the predicted codec representation; Encoder-Mix, which removes the enrollment representation; and Encoder-Ref, which removes the mixture representation. Encoder-Mix performs comparably to Encoder-All, whereas Encoder-Ref shows a clear degradation, particularly in dWER and speaker similarity. These results indicate that mixture-side information is critical in the refinement stage. They also suggest that the encoder-only module functions as more than a post-processing vocoder, as it still depends on task-relevant conditioning for target reconstruction.

\subsubsection{Analysis of the Decoder-Only Language Model}
Table~\ref{tbl:ablation} summarizes the ablation results under different decoder-only input-output configurations. First, varying the number of predicted RVQ layers from one to three leads to only minor differences. This suggests that predicting a small number of coarse codec layers is sufficient for the decoder-only model, while the remaining detail can be recovered by the refinement module.

Second, the Ref output setting, in which the model generates both the enrollment and enhanced segments, performs comparably to the baseline. This result suggests that strict output-length alignment is not critical in the current framework.

Third, replacing continuous acoustic inputs with discrete codec inputs leads to a consistent performance drop (Discrete IO), and WavLM-based~\cite{wavlm} input features further degrade semantic fidelity and speaker similarity. These findings indicate that continuous acoustic conditioning better preserves target-related fine-grained information than discretized or more abstract representations in this task.

\subsection{Ablation Results of USEF-Laura-TSE}

We next analyze USEF-Laura-TSE on Libri2Mix from three perspectives: the effect of the discriminative front-end, the role of SI-SDR regularization, and the trade-off between autoregressive and non-autoregressive inference.

\subsubsection{ Impact of the Discriminative Front-End}

We first examine whether introducing a discriminative front-end improves the overall performance of the generative model. Table~\ref{tbl:discriminative} compares the purely discriminative model USEF-TFGridNet-S, the purely generative model LauraTSE, and the proposed two-stage model USEF-Laura-TSE-S under different training strategies.

% To investigate the effect of introducing a discriminative front-end prior to the generative model, we consider two training strategies: (i) a \emph{frozen} setting, where the pre-trained discriminative front-end is kept fixed and used solely as a feature extractor; and (ii) an \emph{unfrozen} setting, where the discriminative front-end is jointly optimized together with the generative back-end. Table~\ref{tbl:discriminative} reports the test results on Libri2Mix for the purely discriminative model (USEF-TFGridNet-S), the purely generative model (LauraTSE), and the proposed two-stage model (USEF-LauraTSE-S).
%
\begin{table*}
  \caption{Results of the discriminative–generative models on the Libri2Mix clean test set. In the “Category” column, “D” denotes a discriminative model, “G” denotes a generative model, and “D–G” denotes a discriminative–generative model. In the “Training Data” column, “1” indicates training on LibriMix, while “2” denotes training with online mixing on LibriSpeech followed by fine-tuning on Libri2Mix. %USEF-TFGridNet-S refers to the USEF-TFGridNet model with two TF-GridNet blocks, and USEF-Laura-TSE-S denotes the discriminative–generative model that employs USEF-TFGridNet-S as the discriminative front-end and LauraTSE as the generative back-end.
  “SBERT” denotes SpeechBERT score.}
  \begin{center}
    % \setlength{\tabcolsep}{7.5pt}
    % \begin{adjustbox}{width=\linewidth}
    \begin{tabular}{cccccccccccc}
      \Xhline{2\arrayrulewidth}
      \multirow{2}{*}{Model}           & \multirow{2}{*}{Category} & \multirow{2}{*}{Frozen} & \multirow{2}{*}{\makecell{Training \\ Data}} & \multicolumn{3}{c}{DNSMOS $\uparrow$} & \multirow{2}{*}{NISQA $\uparrow$} & \multirow{2}{*}{SBERT $\uparrow$} & \multirow{2}{*}{dWER $\downarrow$} & \multirow{2}{*}{WavLM $\uparrow$} & \multirow{2}{*}{Wespeaker $\uparrow$} \\
                                       &                           &                         &                                & SIG     & BAK     & OVRL   &                        &                             &                       &                        &                                \\ \hline
      USEF-TFGridNet-S                 & D                         & -                       & 1                      & 3.308   & 3.745   & 2.926  & 3.349                  & 0.807                       & 0.228                 & 0.961                  & 0.912                          \\ \hline
      \multirow{2}{*}{LauraTSE}        & \multirow{2}{*}{G}        & \multirow{2}{*}{-}      & 1                      & 3.629   & 4.102   & 3.360  & 4.241                  & 0.879                       & 0.241                 & 0.965                  & 0.847                               \\
                                       &                           &                         & 2                      & 3.609   & 4.084   & 3.336  & 4.333                  & 0.908                       & 0.159                 & 0.974                  & 0.876                          \\ \hline
      \multirow{3}{*}{USEF-Laura-TSE-S} & \multirow{3}{*}{D-G}      & \cmark                     & 1                      & 3.606   & 4.100   & 3.344  & 4.304                  & 0.869                       & 0.266                 & 0.963                  & 0.851                          \\
                                       &                           & \xmark                  & 1                      & 3.609   & 4.086   & 3.341  & 4.350                  & 0.910                       & 0.153                 & 0.973                  & 0.879                          \\
                                       &                           & \xmark                  & 2                      & 3.592   & 4.061   & 3.313  & 4.453                  & 0.925                       & 0.120                 & 0.978                  & 0.895                          \\
                                       \Xhline{2\arrayrulewidth}
      \end{tabular}
    \label{tbl:discriminative}
  % \end{adjustbox}
  \end{center}
  % \vspace{-10pt}
\end{table*}

Compared with LauraTSE, USEF-Laura-TSE-S with joint training achieves improved or comparable performance across multiple metrics. In particular, under the larger-data setting, the dWER is reduced from 0.159 to 0.120, while SpeechBERT and both speaker-similarity metrics are also improved. These results suggest that the discriminative front-end provides more structured and less interference-corrupted intermediate representations, thereby reducing the burden of coarse target alignment for the generative back-end.

We also compare frozen and unfrozen front-end training. Although the frozen setting still improves over the purely generative model in some perceptual metrics, it performs noticeably worse than the jointly trained version in semantic-related metrics such as SpeechBERT and dWER. This indicates that end-to-end optimization is beneficial because it allows the front-end and back-end to adapt to each other.

Compared with the standalone discriminative baseline, the two-stage system consistently improves perceptual quality while retaining strong semantic fidelity and speaker consistency. This supports the main motivation of the proposed framework: the discriminative front-end contributes target controllability and interference suppression, while the generative back-end improves perceptual reconstruction quality.

\begin{table*}[]
  \caption{Results on the Libri2Mix clean test set for the discriminative–generative models with an additional SI-SDR loss. In the “O” column, “D” denotes the output of the discriminative model, and “G” denotes the output of the generative model. %In the “Model” column, USEF-Laura-TSE-S refers to the discriminative–generative system that uses USEF-TFGridNet-S as the discriminative front-end and LauraTSE as the generative back-end, while USEF-Laura-TSE-L employs USEF-TFGridNet-L (with six TF-GridNet blocks) as the discriminative front-end and LauraTSE as the generative back-end. For all USEF-Laura-TSE variants, the discriminative front-end is pre-trained only on Libri2Mix. 
  In the “Training Data” column, “1” indicates training on LibriMix, and “2” denotes training with online mixing on LibriSpeech followed by fine-tuning on Libri2Mix. “SBERT” denotes the SpeechBERT score.}
  \begin{center}
    % \setlength{\tabcolsep}{7.5pt}
    % \begin{adjustbox}{width=\linewidth}
      \begin{tabular}{cccccccccccc}
        \Xhline{2\arrayrulewidth}
        \multirow{2}{*}{Model}               & \multirow{2}{*}{\begin{tabular}[c]{@{}c@{}}Training\\ Data\end{tabular}}                  & \multirow{2}{*}{\begin{tabular}[c]{@{}c@{}}SI-SDR\\Loss?\end{tabular}} & \multirow{2}{*}{O} & \multicolumn{3}{c}{DNSMOS} & \multirow{2}{*}{NISQA} & \multirow{2}{*}{SBERT} & \multirow{2}{*}{dWER} & \multirow{2}{*}{WavLM} & \multirow{2}{*}{Wespeaker} \\
                                          &                   &         &                    & SIG     & BAK     & OVRL   &                        &                        &                       &                        &                     \\ \hline
        % USEF-TFGridNet-S                  & 1                 & -       & D                  &         &         &        &                        &                        &                       &                        &                     \\
        % USEF-TFGridNet-S                  & 2                 & -       & D                  &         &         &        &                        &                        &                       &                        &                     \\
        USEF-TFGridNet-L                  & 1                  & -      & D                  & 3.514   & 4.041   & 3.249  & 4.370                  & 0.909                  & 0.104                 & 0.982                  & 0.953               \\
        USEF-TFGridNet-L                  & 2                  & -      & D                  & 3.555   & 4.051   & 3.272  & 4.319                  & 0.935                  & 0.075                & 0.988                  & 0.968               \\
        LauraTSE                          & 2                  & -      & G                  & 3.609   & 4.066   & 3.336  & 4.333                  & 0.908                  & 0.159                 & 0.974                  & 0.876               \\ \hline
        \multirow{3}{*}{USEF-Laura-TSE-S} & 2                  & No     & D                  & 1.187  & 1.144   & 1.100  & 1.014                  & 0.451                  & 0.693                  & 0.672                  & 0.642               \\
                                          & 2                   & No     & G                  & 3.592   & 4.061   & 3.313  & 4.453                  & 0.925                  & 0.120                 & 0.978                  & 0.895               \\
                                          & 2                   & Yes    & D                  & 3.422   & 3.661   & 2.979  & 3.172                  & 0.884                  & 0.113                 & 0.977                  & 0.937               \\
                                          & 2                   & Yes    & G                  & 3.603   & 4.080   & 3.329  & 4.416                  & 0.915                  & 0.154                 & 0.975                  & 0.880               \\ \hline
        \multirow{2}{*}{USEF-Laura-TSE-L} & 2                   & Yes    & D                  & 3.528   & 3.955   & 3.202  & 3.648                  & 0.933                  & 0.076                 & 0.987                  & 0.950                    \\
                                          & 2                   & Yes    & G                  & 3.592   & 4.075   & 3.319  & 4.450                  & 0.934                  & 0.117                 & 0.982                  & 0.902                    \\ \hline        
      \Xhline{2\arrayrulewidth}
      \end{tabular}
    \label{tbl:dg_sisdr}
  % \end{adjustbox}
  \end{center}
  \vspace{-10pt}
\end{table*}

\subsubsection{Effect of SI-SDR Regularization on the Two-Stage Framework}
We next investigate the effect of SI-SDR regularization by introducing an auxiliary SI-SDR loss on the discriminative front-end. We compare standalone discriminative and generative models as well as different configurations of the proposed two-stage framework under the Training Data = 2 setting.

From Table~\ref{tbl:dg_sisdr}, the purely discriminative model (USEF-TFGridNet-L) achieves strong performance in semantic and speaker-related metrics, including low dWER and high speaker similarity scores, while the generative model (LauraTSE) shows better perceptual quality but higher dWER.

For the proposed two-stage model without SI-SDR loss, joint training leads to a clear separation between the roles of the two modules. The discriminative output shows degraded perceptual quality compared with the pre-trained model, while the generative output improves over LauraTSE in dWER and speaker similarity metrics. This indicates that the discriminative front-end shifts its focus toward producing representations more suitable for the generative back-end.

When SI-SDR regularization is applied, the discriminative output becomes more stable in semantic-related metrics, with reduced dWER and improved speaker similarity compared with the pre-trained discriminative model. However, this is accompanied by a slight degradation in perceptual quality. For the generative output, SI-SDR regularization leads to minor changes in perceptual metrics and a moderate increase in dWER compared with the non-regularized setting.

A similar trend is observed for the larger front-end configuration (USEF-Laura-TSE-L). SI-SDR regularization improves the semantic consistency of the discriminative output, while slightly limiting the improvement of the generative output.

Overall, the results show that SI-SDR regularization improves the stability of the discriminative front-end in terms of semantic-related metrics, but introduces a trade-off in perceptual quality and reduces the flexibility of the generative back-end. This suggests that the strength of the SI-SDR constraint should be carefully balanced in the two-stage framework.
% We next study the effect of SI-SDR regularization on USEF-Laura-TSE by examining how an auxiliary SI-SDR loss on the discriminative front-end affects both the front-end output and the final generative output.

% We first compare the standalone discriminative and generative models under the Training Data = 2 setting. As shown in the first three rows of Table~\ref{tbl:dg_sisdr}, the discriminative model USEF-TFGridNet-L achieves strong overall performance, with a DNSMOS-OVRL score of 3.272, a NISQA score of 4.319, a low dWER of 0.075, and speaker similarity scores exceeding 0.98 for both WavLM and WeSpeaker. In contrast, the purely generative model LauraTSE slightly outperforms USEF-TFGridNet-L in perceptual quality metrics (DNSMOS-OVRL: 3.336; NISQA: 4.333), but exhibits inferior semantic consistency and speaker preservation, as reflected by a higher dWER (0.159) and a lower WeSpeaker score (0.876). A similar trend is observed for USEF-TFGridNet-S, where the discriminative model maintains advantages in dWER and speaker similarity, while lagging behind the generative model in perceptual quality.

%
\begin{table*}[]
  \caption{Results on the Libri2Mix clean test set for USEF-Laura-TSE under auto-regressive (AR) and non-auto-regressive (NAR) inference. %In the “Model” column, “USEF-TFGridNet-L + LauraTSE (split)” denotes that USEF-TFGridNet-L and LauraTSE are trained separately, and during NAR inference the outputs of USEF-TFGridNet-L are used as pseudo labels for the generative model. 
  In the “Inference Mode” column, “AR” indicates autoregressive inference and “NAR” indicates non-autoregressive inference. “R” denotes the injection ratio of the discriminative outputs. “SBERT” denotes the SpeechBERT score.}
  \begin{center}
    % \setlength{\tabcolsep}{7.5pt}
    % \begin{adjustbox}{width=\linewidth}
      \begin{tabular}{ccccccccccccc}
      \Xhline{2\arrayrulewidth}
        \hline
        \multirow{2}{*}{Model}                                                                               & \multirow{2}{*}{\begin{tabular}[c]{@{}c@{}}Inference\\ Mode\end{tabular}} & \multirow{2}{*}{\begin{tabular}[c]{@{}c@{}}SI-SDR\\ Loss?\end{tabular}} & \multirow{2}{*}{O} & \multirow{2}{*}{R} & \multicolumn{3}{c}{DNSMOS} & \multirow{2}{*}{NISQA} & \multirow{2}{*}{SBERT} & \multirow{2}{*}{dWER} & \multirow{2}{*}{WavLM} & \multirow{2}{*}{Wespeaker} \\ \cline{6-8}
                                                                                                          &                                                                  &                                                      &                  &                    & SIG     & BAK     & OVRL   &                        &                        &                       &                        &                     \\ \hline
        \multirow{5}{*}{USEF-Laura-TSE-S}                                                                 & \multirow{2}{*}{AR}                                              & No                                                    & D               & -                   & 1.187  & 1.144   & 1.100  & 1.014                  & 0.451                  & 0.693                  & 0.672                  & 0.642                    \\
                                                                                                          &                                                                  & No                                                    & G               & -                   & 3.592   & 4.061   & 3.313  & 4.453                  & 0.925                  & 0.120                 & 0.978                  & 0.895               \\ \cline{2-13}
                                                                                                          & \multirow{3}{*}{NAR}                                             & No                                                    & G               & 0.0                & 2.647   & 2.061   & 1.905  & 1.864                  & 0.473                  & 1.024                 & 0.796                  & 0.661               \\
                                                                                                          &                                                                  & No                                                    & G               & 0.5                & 2.452   & 1.768   & 1.724  & 1.635                  & 0.454                  & 1.069                 & 0.782                  & 0.647               \\
                                                                                                          &                                                                  & No                                                    & G               & 1.0                & 1.844   & 1.353   & 1.404  & 1.362                  & 0.424                  & 1.105                 & 0.752                  & 0.637               \\ \hline
        \multirow{5}{*}{USEF-Laura-TSE-S}                                                                 & \multirow{2}{*}{AR}                                              & Yes                                                    & D               & -                  & 3.422   & 3.661   & 2.979  & 3.172                  & 0.884                  & 0.113                 & 0.977                  & 0.934               \\
                                                                                                          &                                                                  & Yes                                                    & G               & -                  & 3.603   & 4.080   & 3.329  & 4.416                  & 0.915                  & 0.154                 & 0.975                  & 0.880               \\ \cline{2-13}
                                                                                                          & \multirow{3}{*}{NAR}                                             & Yes                                                    & G               & 0.0                & 3.590   & 4.027   & 3.291  & 4.217                  & 0.910                  & 0.149                 & 0.975                  & 0.881               \\
                                                                                                          &                                                                  & Yes                                                    & G               & 0.5                & 3.578   & 3.991   & 3.263  & 4.099                  & 0.907                  & 0.148                 & 0.975                  & 0.882               \\
                                                                                                          &                                                                  & Yes                                                    & G               & 1.0                & 3.568   & 3.944   & 3.232  & 3.960                  & 0.906                  & 0.133                 & 0.975                  & 0.883               \\ \hline
        \multirow{5}{*}{\begin{tabular}[c]{@{}c@{}}USEF-TFGridNet-L\\ \\ + LauraTSE (split)\end{tabular}} & \multirow{2}{*}{AR}                                              & -                                                    & D               & -                  & 3.555   & 4.051   & 3.272  & 4.319                  & 0.935                  & 0.075                & 0.988                  & 0.968               \\
                                                                                                          &                                                                  & -                                                    & G                & -                  & 3.609   & 4.084   & 3.336  & 4.333                  & 0.908                  & 0.159                 & 0.974                  & 0.876               \\ \cline{2-13}
                                                                                                          & \multirow{3}{*}{NAR}                                             & -                                                    & G                & 0.0                & 3.587   & 4.089   & 3.322  & 4.512                  & 0.881                  & 0.216                 & 0.969                  & 0.866               \\
                                                                                                          &                                                                  & -                                                    & G                & 0.5                & 3.604   & 4.101   & 3.343  & 4.553                  & 0.898                  & 0.166                 & 0.972                  & 0.872               \\
                                                                                                          &                                                                  & -                                                    & G                & 1.0                & 3.619   & 4.114   & 3.363  & 4.583                  & 0.913                  & 0.120                 & 0.974                  & 0.878               \\ \hline
        \multirow{5}{*}{USEF-Laura-TSE-L}                                                                 & \multirow{2}{*}{AR}                                              & Yes                                                    & D               & -                  & 3.528   & 3.955   & 3.202  & 3.648                  & 0.933                  & 0.076                & 0.987                  & 0.950               \\
                                                                                                          &                                                                  & Yes                                                    & G               & -                  & 3.592   & 4.075   & 3.319  & 4.450                  & 0.934                  & 0.117                 & 0.982                  & 0.902               \\ \cline{2-13}
                                                                                                          & \multirow{3}{*}{NAR}                                             & Yes                                                    & G               & 0.0                & 3.580   & 4.048   & 3.294  & 4.346                  & 0.927                  & 0.115                 & 0.981                  & 0.901               \\
                                                                                                          &                                                                  & Yes                                                    & G               & 0.5                & 3.574   & 4.035   & 3.283  & 4.316                  & 0.927                  & 0.112                 & 0.981                  & 0.902               \\
                                                                                                          &                                                                  & Yes                                                    & G               & 1.0                & 3.570   & 4.022   & 3.272  & 4.302                  & 0.929                  & 0.099                 & 0.982                  & 0.903               \\
\Xhline{2\arrayrulewidth}
      \end{tabular}
    \label{tbl:dg_nar}
  % \end{adjustbox}
  \end{center}
  % \vspace{-10pt}
\end{table*}

\begin{table*}
  \caption{Results on Libri2Mix clean.  In the "Category" column, "G" refers to generative models, while "D" refers to discriminative models.}
  \renewcommand{\arraystretch}{1.2}
  \begin{center}
    \setlength{\tabcolsep}{7.5pt}
    \begin{tabular}{ccccccccccc}
      \Xhline{2\arrayrulewidth}
      \multirow{2}{*}{Model} & \multirow{2}{*}{Category} & \multicolumn{3}{c}{DNSMOS $\uparrow$} & \multirow{2}{*}{NISQA $\uparrow$} & \multirow{2}{*}{SBERT $\uparrow$} & \multirow{2}{*}{dWER $\downarrow$} & \multirow{2}{*}{WavLM $\uparrow$} & \multirow{2}{*}{Wespeaker $\uparrow$} \\
      \cline{3-5}
                             &                           & SIG & BAK & OVL &                         &                      &                         &                          &                          \\
      \hline
      Mixture                & -                         & 3.383 & 3.098 & 2.653 & 2.453 & 0.572 & 0.792 & 0.847 & 0.759 \\
      \hline
      Spex+~\cite{ge2020spex+}  & D                         & 3.472  & 4.027  & 3.186  & 3.349 & 0.878 & 0.148 & 0.973 & 0.935 \\
      WeSep~\cite{wang2024wesep} & D                      & 3.486 & 3.838 & 3.118 & 3.892 & 0.895 & 0.123     & 0.980 & 0.945 \\
      USEF-TFGridNet-L~\cite{11012711} & D                      & 3.555 & 4.051 & 3.272 & 4.319 & \textbf{0.935} & \textbf{0.0747}     & \textbf{0.988} & \textbf{0.968} \\
      \hline
      TSELM-L \cite{tang2024tselm} & G                   & 3.489 & 4.041 & 3.212 & 3.961 & 0.793 & 0.297 & 0.887 & 0.627 \\
      AnyEnhance \cite{zhang2025anyenhance} & G          & \textbf{3.638} & 4.066 & \textbf{3.353} & 4.277 & 0.735 & -     & 0.914 & - \\
      \hline
      LauraTSE               & G                         & 3.609 & \textbf{4.084} & 3.336 & 4.333 & 0.908 & 0.159 & 0.974 & 0.876 \\
      USEF-Laura-TSE-L & D-G & 3.592 & 4.075 & 3.319 & \textbf{4.450} & 0.934 & 0.117 & 0.982 & 0.902 \\
      \Xhline{2\arrayrulewidth}
    \end{tabular}
    \label{tbl:libri2mix_main_exp}
  \end{center}
  \vspace{-10pt}
\end{table*}

\subsubsection{Auto-Regressive and Non-Auto-Regressive Inference Strategies}

% Under identical training conditions, the proposed discriminative-generative framework supports two inference modes at test time: auto-regressive (AR) and non-auto-regressive (NAR). In the standard setting, the decoder-only language model generates discrete target speech representations in an auto-regressive, frame-by-frame manner. Within the discriminative--generative architecture, an alternative NAR inference strategy can be adopted by leveraging the outputs of the discriminative front-end as pseudo labels for the generative model. Specifically, in the NAR inference mode, the training procedure remains unchanged. At inference time, the target speech representations produced by the discriminative front-end are used as pseudo-labels and injected into the decoder-only language model's decoding sequence, along with the mixed and enrollment speech representations. By controlling the pseudo-label injection ratio $R$, the inference process can be flexibly adjusted between fully generative decoding ($R=0$) and firm reliance on the discriminative front-end estimates ($R=1$). A smaller $R$ preserves more generative flexibility, whereas a larger $R$ emphasizes the robustness and controllability of the discriminative front-end. Comparative experiments are conducted on four configurations: USEF-LauraTSE-S with a frozen front-end, USEF-LauraTSE-S with an additional SI-SDR loss, a decoupled-training discriminative--generative model, and USEF-LauraTSE-S with SI-SDR loss under decoupled training. The results are reported in Table~\ref{tbl:dg_nar}.
Under identical training conditions, the proposed discriminative–generative framework supports both auto-regressive (AR) and non-auto-regressive (NAR) inference. In the AR setting, the decoder-only model generates target speech representations sequentially. In contrast, NAR inference leverages the outputs of the discriminative front-end as auxiliary guidance, controlled by an injection ratio R, where R=0 corresponds to pure generative decoding and R=1 fully relies on the discriminative output.

Table~\ref{tbl:dg_nar} reports results under different configurations. Overall, AR inference achieves the best perceptual quality for both USEF-LauraTSE-S and USEF-LauraTSE-L, as reflected by higher DNSMOS and NISQA scores. However, AR inference also yields higher dWER compared with NAR settings, indicating weaker semantic stability.

For NAR inference, increasing the injection ratio R consistently leads to a reduction in dWER while slightly degrading perceptual metrics such as DNSMOS and NISQA. For example, in USEF-LauraTSE-S with SI-SDR loss, dWER decreases from 0.154 to 0.133 as R increases from 0 to 1, while DNSMOS-OVRL and NISQA show a moderate decrease. A similar trend is observed for USEF-LauraTSE-L, where higher R improves dWER (0.112 to 0.099) but slightly reduces perceptual quality.

The cascaded baseline (USEF-TFGridNet-L + LauraTSE, split training) achieves the highest perceptual scores (DNSMOS-OVRL and NISQA) among all configurations. However, it performs worse in dWER compared with the jointly trained USEF-LauraTSE models, indicating inferior semantic consistency despite better perceptual quality.

Overall, the results indicate a trade-off between perceptual quality and semantic stability. AR inference favors perceptual quality, while NAR inference improves intelligibility and semantic consistency by incorporating discriminative guidance through the injection ratio R.

\begin{table*}
\caption{The results on validation set of Urgent Challenge for BSRNN-Laura-SE.}
\begin{center}
\begin{tabular}{c|c|ccccccccc}
\Xhline{2\arrayrulewidth}
\textbf{Model}                         & \textbf{Codec}            & \textbf{\begin{tabular}[c]{@{}c@{}}Inference\\ Mode\end{tabular}} & \textbf{R} & \textbf{OVEL}  & \textbf{NISQA} & \textbf{SCOREQ} & \textbf{PESQ}  & \textbf{ESTOI} & \textbf{SPK\_SIM} & \textbf{cAcc}  \\ \hline
\multirow{4}{*}{BSRNN-Laura-SE}        & \multirow{4}{*}{FunCodec} & AR                                                                & -          & 3.344          & 4.013          & 3.974           & 1.624          & 0.510          & 0.545             & 0.642          \\
                                       &                           & \multirow{3}{*}{NAR}                                              & 0          & 3.230          & 3.600          & 3.518           & 1.703          & 0.706          & 0.679             & 0.728          \\
                                       &                           &                                                                   & 0.5        & 3.205          & 3.512          & 3.433           & 1.742          & 0.722          & 0.682             & 0.776          \\
                                       &                           &                                                                   & 1.0        & 3.181          & 3.431          & 3.371           & 1.788          & 0.735          & 0.684             & 0.806          \\ \hline
\multirow{4}{*}{BSRNN-Laura-SE(split)} & \multirow{4}{*}{FunCodec} & AR                                                                & -          & \textbf{3.409} & \textbf{4.293} & \textbf{4.269}  & 1.824          & 0.749          & 0.643             & 0.829          \\
                                       &                           & \multirow{3}{*}{NAR}                                              & 0          & 3.372          & 3.996          & 4.062           & 1.942          & 0.777          & 0.665             & 0.845          \\
                                       &                           &                                                                   & 0.5        & 3.369          & 3.999          & 4.060           & 1.944          & 0.777          & 0.665             & 0.845          \\
                                       &                           &                                                                   & 1.0        & 3.359          & 3.902          & 3.990           & \textbf{1.978} & \textbf{0.781} & \textbf{0.671}    & \textbf{0.847} \\ \hline
\multirow{4}{*}{BSRNN-Laura-SE(split)} & \multirow{4}{*}{FlowDec}  & AR                                                                & -          & 3.256          & 3.785          & 3.409           & 1.762          & 0.751          & 0.678             & 0.841          \\
                                       &                           & \multirow{3}{*}{NAR}                                              & 0          & 3.216          & 3.688          & 3.380           & 1.785          & 0.765          & 0.664             & 0.843          \\
                                       &                           &                                                                   & 0.5        & 3.216          & 3.695          & 3.357           & 1.807          & 0.766          & 0.670             & 0.845          \\
                                       &                           &                                                                   & 1.0        & 3.222          & 3.722          & 3.353           & 1.833          & 0.768          & 0.674             & 0.847          \\ \Xhline{2\arrayrulewidth}
\end{tabular}
\label{tbl:urgent-abl}
\end{center}
\end{table*}

\begin{table*}
\caption{The results of BSRNN-Laura-SE and other SE models on the validation set of the Urgent Challenge.}
\begin{center}
\begin{tabular}{c|c|ccccccc}
\Xhline{2\arrayrulewidth}
\textbf{Model} & \textbf{Codec} & \textbf{OVEL} & \textbf{NISQA} & \textbf{SCOREQ} & \textbf{PESQ} & \textbf{ESTOI} & \textbf{SPK\_SIM} & \textbf{cAcc} \\ \hline
BSRNN          & -              & 3.13          & 3.48           & 3.32            & 2.57          & 0.85           & 0.78              & 0.875         \\
BSRNN-FLOW     & -              & 3.19          & 3.83           & 3.35            & 2.15          & 0.81           & 0.74              & 0.809         \\ \hline
subatomicseer  & -              & 3.17          & 3.72           & 3.56            & 2.68          & 0.85           & 0.80              & 0.880         \\
GHW            & -              & 3.20          & 4.08           & 3.77            & 2.74          & 0.85           & 0.79              & 0.854         \\
baird          & -              & 3.13          & 3.57           & 3.43            & 2.61          & 0.85           & 0.79              & 0.877         \\ \hline
BSRNN-Laura-SE(split) & FunCodec       & 3.36          & 3.90           & 3.99            & 1.98          & 0.78           & 0.67              & 0.847         \\
% BSRNN-Laura-SE & FlowDec        &               &                &                 &               &                &                   &               \\ 
\Xhline{2\arrayrulewidth}
 \end{tabular}
 \label{tbl:urgent-cp}
 \end{center}
 \vspace{-10pt}
\end{table*}

\subsection{Comparison With Previous TSE Models}
Table~\ref{tbl:libri2mix_main_exp} summarizes the overall experimental results on the Libri2Mix dataset. The proposed discriminative–generative model USEF-Laura-TSE-L achieves a more balanced performance across perceptual quality, semantic fidelity, and speaker consistency. Compared with the purely generative LauraTSE, USEF-Laura-TSE-L significantly improves semantic consistency and speaker similarity, with dWER reduced from 0.159 to 0.117, and speaker similarity increased from 0.974/0.876 to 0.982/0.902 (WavLM/WeSpeaker), while maintaining comparable DNSMOS-OVRL and achieving the best NISQA score (4.450) among all systems. This indicates that a stronger discriminative front-end (USEF-TFGridNet-L) provides more reliable and structured intermediate representations, which effectively guide the generative back-end toward improved content stability without sacrificing perceptual quality.

Compared with the strong discriminative baseline USEF-TFGridNet-L~\cite{11012711}, which attains the best dWER and speaker similarity, USEF-Laura-TSE-L substantially improves perceptual quality (DNSMOS-OVRL and NISQA), demonstrating that the discriminative–generative framework effectively bridges the gap between discriminative robustness and generative naturalness. Moreover, despite being trained on only 460 hours of data, USEF-Laura-TSE-L achieves performance comparable to or better than large-scale generative systems such as AnyEnhance~\cite{zhang2025anyenhance}, highlighting the data efficiency and effectiveness of task-oriented discriminative–generative modeling for target speaker extraction.

Overall, these results confirm that combining a strong discriminative front-end with a generative AR decoder-only back-end yields a robust and well-balanced solution, validating the effectiveness of the proposed discriminative–generative two-stage paradigm.

\subsection{Ablation study of BSRNN-Laura-SE}

Table~\ref{tbl:urgent-abl} reports the validation results of BSRNN-Laura-SE on the URGENT Challenge. The results show that the proposed discriminative–generative two-stage framework can be extended from target speaker extraction to speech enhancement. In the split setting with FunCodec, the model achieves the best OVEL, NISQA, and SCOREQ scores under AR inference, while NAR inference with R=1.0 obtains the best PESQ, ESTOI, SPK\_SIM, and cAcc. This trend is consistent with the observations in USEF-Laura-TSE: AR inference tends to benefit overall perceptual quality, whereas NAR inference with more front-end predictions provides more stable reconstruction and improves intelligibility- and content-related metrics.

The split training strategy performs better than the jointly trained BSRNN-Laura-SE in most metrics. This may indicate that, for speech enhancement, fixing a pre-trained BSRNN front-end provides a more stable input distribution for the generative back-end. This differs from USEF-Laura-TSE, where SI-SDR regularization helps stabilize the TSE front-end. A possible reason is that SE does not involve target-speaker selection, and the BSRNN front-end can already produce a reasonable enhanced estimate. Thus, an SI-SNR constraint alone may not be sufficient for effective joint optimization of waveform fidelity, spectral structure, and perceptual quality.

FunCodec achieves better results than FlowDec in most metrics, although FlowDec operates at 48 kHz. This may be related to both the evaluation protocol and the SE task itself. The adopted metrics mainly reflect intelligibility, perceptual quality, and distortion in speech-relevant frequency bands, while the benefit of high-frequency modeling may not be fully captured. In addition, high-frequency details in noisy conditions are harder to predict reliably and may introduce artifacts. These results suggest that the two-stage framework can be applied to both TSE and SE, with broadly similar AR/NAR inference trends across the two tasks.

% \usepackage{multirow}

\subsection{Comparison study of BSRNN-Laura-SE}
Table~\ref{tbl:urgent-cp} compares BSRNN-Laura-SE with the top-ranked systems on the URGENT Challenge validation set. BSRNN-Laura-SE achieves the best OVEL and SCOREQ scores among all compared systems, indicating competitive overall perceptual quality and speech quality estimation performance. It also obtains the second-best NISQA score, slightly lower than GHW. These results suggest that the proposed two-stage discriminative–generative framework is effective for improving perceptual quality in the SE task.

However, BSRNN-Laura-SE does not achieve the best results on PESQ, ESTOI, SPK\_SIM, and cAcc. In particular, compared with the top systems, its ESTOI and SPK\_SIM are lower, suggesting that there is still room for improvement in intelligibility preservation and speaker similarity. This may be related to the generative back-end, which tends to improve perceptual quality but may introduce reconstruction deviations that are not always favored by intrusive or similarity-based metrics. Overall, the comparison shows that BSRNN-Laura-SE is competitive with the top-ranked systems, especially in perceptual quality-related metrics, while further optimization is needed for intelligibility and speaker consistency.

\section{Conclusion}

This paper first proposes LauraTSE, a generative target speaker extraction (TSE) method based on an auto-regressive decoder-only language model. By leveraging continuous acoustic features and a neural audio codec, LauraTSE enables end-to-end generative TSE without relying on explicit speaker embeddings. Experimental results demonstrate competitive performance in speech quality, speaker similarity, and semantic consistency, while data-scaling experiments indicate its potential scalability compared with conventional discriminative models. Further analysis shows that coarse auto-regressive generation alone is insufficient for fine-grained acoustic reconstruction, motivating the introduction of an encoder-only LM to refine acoustic details. Building on LauraTSE, this work further presents a discriminative–generative framework, where a USEF-TFGridNet-based front-end provides structured target-related representations to guide generative reconstruction. The results show that this design improves speaker consistency and intelligibility over the purely generative model, suggesting the complementary roles of discriminative and generative modeling in TSE. Additional studies on SI-SDR-constrained training and non-autoregressive inference further demonstrate a controllable trade-off between perceptual quality and semantic robustness. Finally, preliminary results on speech enhancement with BSRNN-Laura-SE suggest that the proposed two-stage paradigm can also be extended beyond TSE, achieving competitive performance on perceptual quality-related metrics in the URGENT Challenge validation set.

\bibliographystyle{IEEEtran}
\bibliography{refs}

\end{document}